\documentclass[conference]{IEEEtran}
\IEEEoverridecommandlockouts
\usepackage{cite}
\usepackage{bm}
\usepackage{amsmath,amssymb,amsfonts}
\usepackage[ruled,vlined,onelanguage]{algorithm2e}
\usepackage{graphicx}
\usepackage{textcomp}
\usepackage{xcolor}
\def\BibTeX{{\rm B\kern-.05em{\sc i\kern-.025em b}\kern-.08em
    T\kern-.1667em\lower.7ex\hbox{E}\kern-.125emX}}
\begin{document}

\newcommand{\cristobal}[1]{\textbf{\textcolor{magenta}{\{Cristobal: #1\}}}}
\newcommand{\rodrigo}[1]{\textbf{\textcolor{blue}{\{Rodrigo: #1\}}}}

\title{GPU Voronoi Diagrams for Random Moving Seeds\\
\thanks{Research supported by ANID Fondecyt grant \#1221357.}
}

\author{\IEEEauthorblockN{Rodrigo Stevenson-Regla}
\IEEEauthorblockA{\textit{Instituto de Informática} \\
\textit{Universidad Austral de Chile}\\
Valdivia, Chile \\
rstevenson@inf.uach.cl}
\and
\IEEEauthorblockN{Crist\'obal A. Navarro}
\IEEEauthorblockA{\textit{Instituto de Informática} \\
\textit{Universidad Austral de Chile}\\
Valdivia, Chile \\
cristobal.navarro@uach.cl}
}
\maketitle

\begin{abstract}
The Voronoi Diagram is a geometrical structure that is widely used in scientific or technological applications where proximity is a relevant aspect to consider, and it also resembles natural phenomena such as cellular banks, rock formations or bee hives, among others. Typically, computing the Voronoi Diagram is done in a static context, that is, the location of the input seeds is defined once and does not change. In this work we study the dynamic case where seeds move, which leads to a dynamic Voronoi Diagram that changes over time. In particular, we consider uniform random moving seeds, for which we propose the \textit{dynamic Jump Flooding Algorithm} (dJFA), a variant of JFA that uses less iterations than the standard JFA. An experimental evaluation shows that dJFA achieves a speedup of up to $\sim 5.3 \times$ over JFA, while maintaining a similarity of at least $88\%$ and close to $100\%$ in many cases. These results contribute with a step towards the achievement of real-time GPU-based computation of dynamic Voronoi diagrams for any particle simulation. 
\end{abstract}


\section{Introduction}
The Voronoi Diagram (VD) is one of the essential structures for computational geometry, along with the convex hull and the Delaunay triangulation which is its dual. The VD provides proximity information of the input seeds (points), something that is required by several scientific and technological applications \cite{Qi2019GPredicatesGI,articleBaTo,10.1007/978-3-642-83539-1_3}.
GPU-based techniques exist and employ a data-parallel design in order to generate the VD efficiently. One of the most known algorithms is the Jump Flooding Algorithm (JFA) \cite{RongJFA,4276119} which is considered one of the fastest VD building techniques. Another efficient approach is the Facet-JFA \cite{10.1145/2683483.2683503} which for some cases is faster than JFA.

The aforementioned approaches are normally used in a static context of seeds, i.e., to have a pixel grid and a set of seeds with fixed locations. If the particles move slowly over time, one could still use any of these state of the art approaches at each time-step, however this would lead to a computational cost that is much higher than what may be really needed, as each state could only be a small displacement from the previous one. Dynamic Voronoi diagrams open the possibility to research on ways to take advantage of the previous state as well as the particles behavior. This work focuses on this research opportunity, by first studying the current GPU rasterized techniques that exist to compute the VD, and then by proposing an algorithm that solves the dynamic case with uniform random moving particles in 2D. Lastly, the proposed method is compared in terms of GPU performance and similarity. 

The remaining Sections cover background on Voronoi Diagrams (Section \ref{sec:voronoi-diagrams}) and the Jump Flooding Algorithm (Section \ref{sec:jfa}), problem statement (Section \ref{sec:problem-statement}), proposed algorithm (Section \ref{sec:proposed-algorithm}), experimental evaluation (Section \ref{sec:experimental-evaluation}) and Conclusions (Section \ref{sec:conclusions}).

\section{Background on Voronoi Diagrams}
\label{sec:voronoi-diagrams}
\begin{figure}[ht!]
\centering
\includegraphics[scale=0.2]{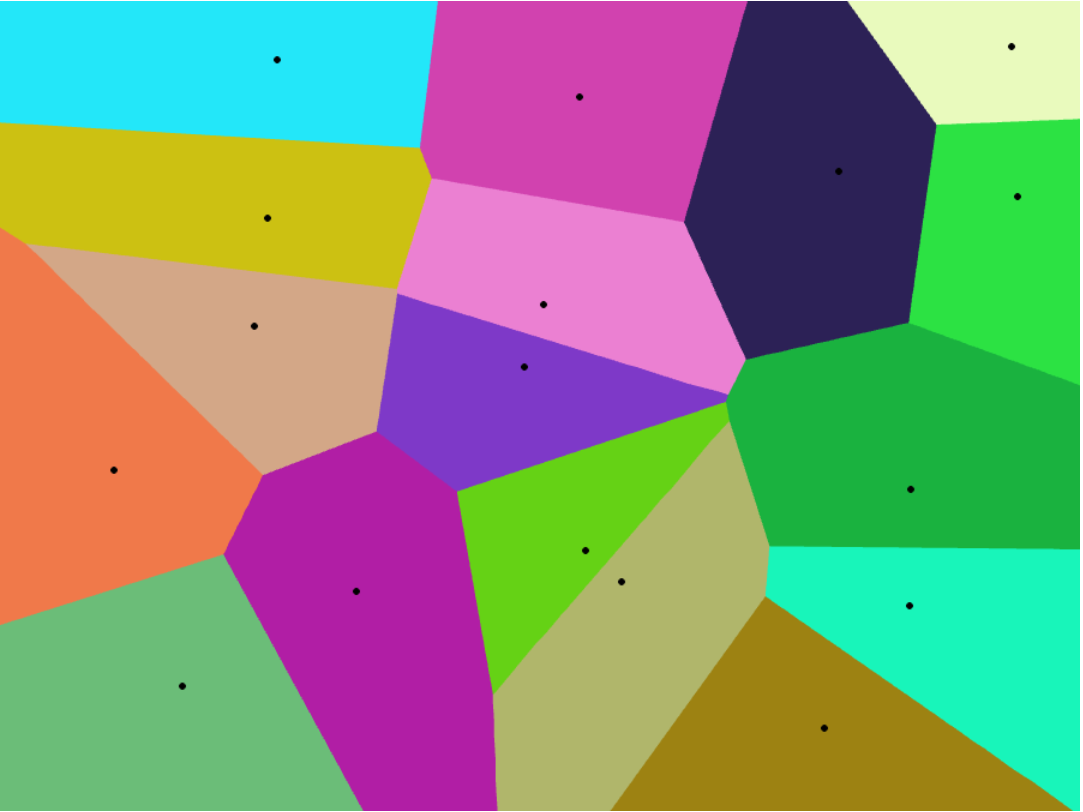}
\caption{An example Voronoi Diagram for 17 seeds in the plane.}
\label{fig:vd-example}
\end{figure}

The Voronoi Diagram (VD) is a geometric structure that  partition the Euclidean space. The resulting structure provides proximity information, where each region surrounding a seed is the space for which all points are closer to that seed than any other, and each frontier where points are equidistant to the two seeds that generate such adjacent regions (Figure \ref{fig:vd-example}).

Voronoi Diagrams define the following parameters:
\begin{itemize}
    \item $X$: Metric space or Grid.
    \item $S$: Set of seeds where $S = \{P_1, P_2, \cdots, P_s\}$.
    \item $R_k$: Voronoi region associated to seed $P_k$.
    \item $d$: Distance function.
\end{itemize}
where regions generate by satisfying the following condition:
\begin{equation}
\label{eq:vd-definition}
    R_k = \{ x \in X | d(x,P_k) \le d(x,P_j) \forall k \neq j,\  P_k,P_j\in S\}
\end{equation}
Eq. (\ref{eq:vd-definition}) tells that the $x$ locations that belong to a region $R_k$ are closer to $P_k$ than any other seed. 

Voronoi diagrams can be used to simulate the structure and dynamics of cell groups \cite{10.1007/978-3-642-29280-4_21,Indermitte:118485} and crystalline compounds \cite{SBDMul,KOBAYASHI2002681}, also they can be applied to solve neighborhood problems related to building roads \cite{4058742,4459318}, among many other applications. One of the main reasons for its use it that the VD can reproduce the formation of natural structures which are of interest in several scientific and technological fields. For the purposes of this research, we have focused on a uniform distribution for the seeds with uniform random movements in 2D. This model, although simple and synthetic, still relates to some existing particle motion models under study \cite{CERDA20188, carter2018gpu}.

\section{Revisiting the Jump Flooding Algorithm (JFA)}
\label{sec:jfa}
\begin{figure}[!ht]
\centering
\includegraphics[scale=0.22]{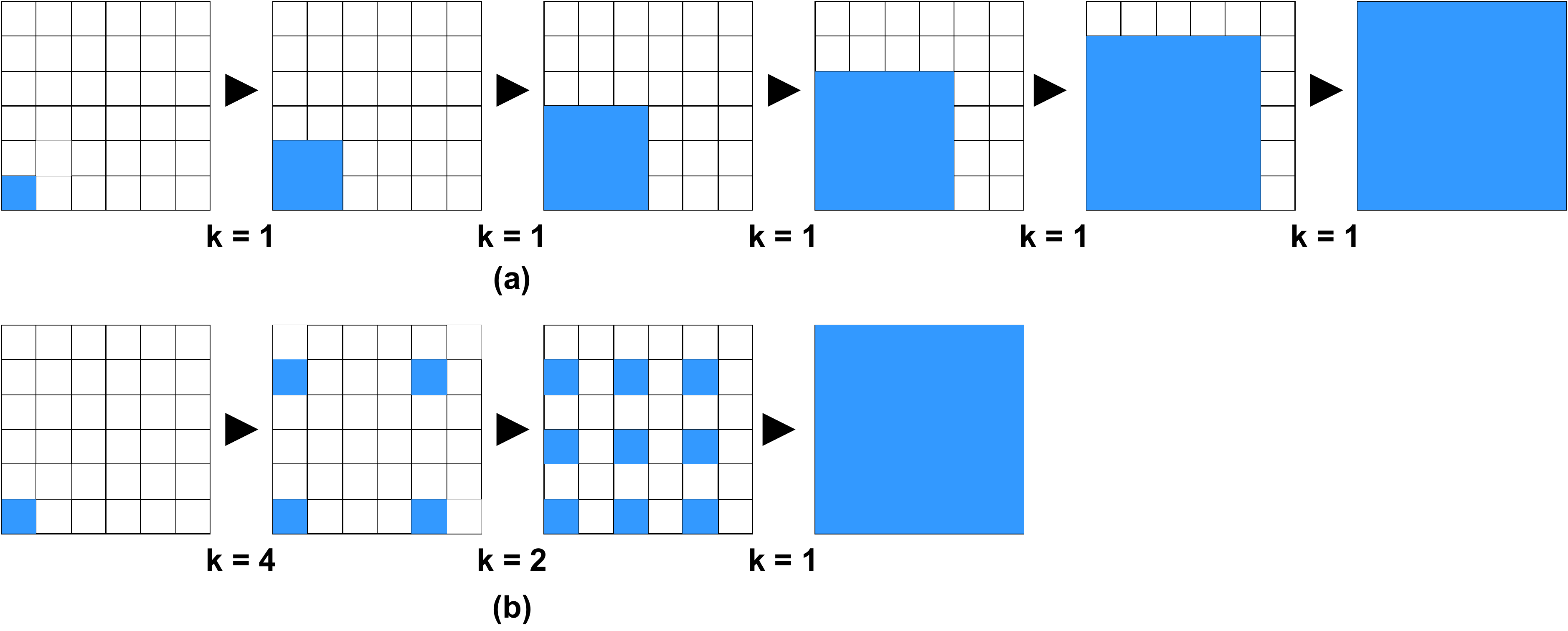}
\caption{(a) Standard flooding algorithm, (b) Jump flooding algorithm.}
\label{fig:flood-comparison}
\end{figure}

The Jump Flooding Algorithm, or simply JFA, was proposed by Rong \& Tan in 2006 \cite{RongJFA,4276119}, as a way to improve the Standard Flooding (StF) method which was one of the best known techniques for constructing the VD with GPU computing. The main problem of StF was that it could not exploit enough parallelism in its first iterations, as the flood was still small. StF works by defining the positions as the starting points for each flood. Then StF floods all neighbors that have a Chebyshev distance of 1 from the existing floods, in parallel, until the grid is fully flooded. Figure \ref{fig:flood-comparison} a) illustrates the process for one seed.

As can be seen, StF fulfills its purpose in 5 iterations, exhibiting a small amount of parallelism at each iteration, as the neighborhood jump step is fixed at $k=1$ throughout all the process.  In comparison, JFA proposes a different type of flooding, by jumping to the nearest neighbors that are at a distance $k_i(n)$ which is defined as a function of the grid size ($n\times n$) and the actual iteration ($i$), i.e.,
\begin{equation}
\label{eq:k_def}
    k_i(n) = \frac{2^{\big\lceil\log_2(N)\big\rceil - 1}}{2^{i-1}}
\end{equation}
The process starts at $i=1$ and terminates with $k_i=1$ (inclusive). Overall, JFA does $\log_2(k) + 1$ steps. As an example, for a $8 \times 8$ grid, there are three iterations with the values of $\{k_1, k_2, k_3\} =\{4,2,1\}$.

The positions for each neighbor are defined in Table \ref{tab:neighbors}. 
\begin{table}[ht!]
\caption{Neighborhood estimation of JFA 2D.}\label{tab:neighbors}
\begin{center}
\resizebox{0.85\columnwidth}{!}{
\begin{tabular}{|l|l|}
\hline
{\bfseries Neighbor} &  {\bfseries Position}\\
\hline
1 &  (\(x_1\),\(y_1\)) = (\(x_0\) + k, \(y_0\))\\
\hline
2 &  (\(x_2\),\(y_2\)) = (\(x_0\) + k, \(y_0\) + k)\\
\hline
3 &  (\(x_3\),\(y_3\)) = (\(x_0\), \(y_0\) + k)\\
\hline
4 &  (\(x_4\),\(y_4\)) = (\(x_0\) - k, \(y_0\) + k)\\
\hline
5 &  (\(x_5\),\(y_5\)) = (\(x_0\) - k, \(y_0\)\\
\hline
6 &  (\(x_6\), \(y_6\)) = (\(x_0\) - k, \(y_0\) - k)\\
\hline
7 &  (\(x_7\), \(y_7\)) = (\(x_0\), \(y_0\) - k)\\
\hline
8 &  (\(x_8\), \(y_8\)) = (\(x_0\) + k, \(y_0\) - k)\\
\hline
\end{tabular}
}
\end{center}
\end{table}
The Table shows that the locations of the neighbors follow a Moore-like neighborhood but with distance $k_i$, and at every iteration of JFA this neighborhood has a shorter distance, due to the reduction of $k$, which reflects the way that JFA propagates in the entire domain as shown in Figure \ref{fig:flood-comparison} b). The performance advantage of JFA comes from the fact that it is capable of doing more parallel work at each iteration, leading to less iterations than StF (Figure \ref{fig:flood-comparison}). It is also worth mentioning that if a pixel wants to propagate to another one that has already been claimed, the distance function is used as a criterion to check which flood carries the closest seed.

JFA has proven to be efficient but is not free of visual errors as stated by Rong \& Tan \cite{4276119}. Due to the distance function, it is possible to take a pixel from a region without claiming another one from his neighborhood at Chebyshev's distance 1. Fortunately this problem can be handled by adding one or two extra rounds of the JFA algorithm (also known as JFA+1 and JFA+2). 
When switching to moving particles, doing JFA at each time-step would no longer be the most efficient method, 

\section{Problem Statement: Dynamic Moving Particles}
\label{sec:problem-statement}
The standard JFA is typically considered in a static context, i.e., there is a defined grid with a fixed seed set and the VD is computed once. However, in a dynamic context where seeds move over time, such as in particle simulations, these can exhibit a behavior where $VD_{t-1}$ and $VD_t$ end up being very similar. In such cases, a direct application of the JFA to each time-step would not be the most efficient approach as it would not be taking advantage of what was computed at the previous time-step, neither considering the movement behavior of particles to see if the $k$ values may have an upper bound smaller than in the standard JFA.
Taking advantage of these properties could save some iterations of the JFA, which would translate into a performance acceleration. 

In this work we consider the case study where particles exhibit a uniform random movement, for which we propose the \textit{dynamic JFA} (dJFA).

\section{A New Dynamic Jump Flooding Algorithm}
\label{sec:proposed-algorithm}
We propose the dynamic Jump Flooding Algorithm (dJFA), which is a modified version of the standard JFA, because it reuses the previous state $VD_{t-1}$ and also redefines the $k_i$ parameter as $\delta_i$ using other considerations. We also consider different types of neighborhood; Moore vs Von Neumann, as well as different distance functions; Euclidean vs Manhattan.

\subsection{Defining the dynamic $\delta_i$ parameter}
\begin{figure}[ht!]
\centering
  \centering
  \includegraphics[scale=0.6]{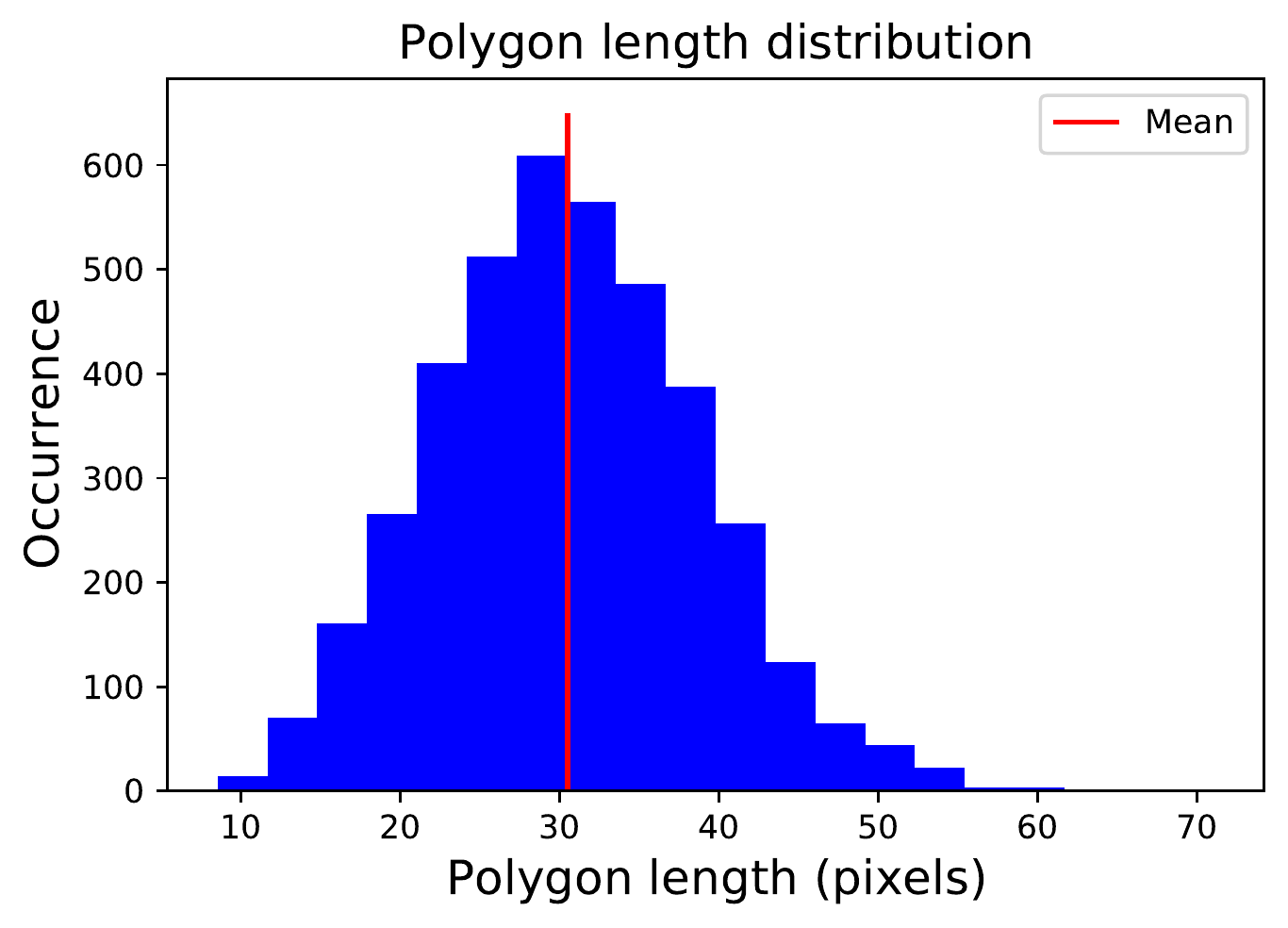}
  \caption{The polygon length histogram resembles a normal distribution.}
  \label{fig:ref-distribution}
\end{figure}

We recall that in JFA defines $k_i$ in terms of the grid size and the current iteration (see Eq.(\ref{eq:k_def})), which leads to a total of $\log_2(k_1)+1$ sequential steps. Here, we aim to redefine $k_i$ as a smaller value in order to produce a smaller number of sequential steps. This new dynamic $k_i$, now named $\delta_i$, takes advantage of the fact that if all particles are uniformly distributed, and move randomly with a uniform distribution, then the $\delta_1$ value does need to begin as large as in Eq. (\ref{eq:k_def}). Moreover, if the density of seeds is high (and uniform by the distribution assumption) then it would be possible to reduce the total amount of generational steps. In the worst case, if the density is too low, it would perform as fast as JFA. Considering that the seeds are randomly distributed with a uniform distribution, one parameter of interest is the average polygon length, which given the assumptions for the seeds, can be approximated to  
\begin{align}
    \label{eq:L_def}
    L_{avg} & \sim \sqrt{\frac{n\cdot n}{s}}.
\end{align}
Figure \ref{fig:ref-distribution} shows the distribution of the polygon (region) lengths for an example set of seeds following a random uniform distribution. 
The average polygon length approximates to a normal distribution, which is a convenient starting point for defining the $\delta_i$ parameter and its limits. Considering that $L$ follows a normal distribution, it is necessary to have a high confidence, such as $99\%$. One approach for this is to consider $2\times L_{avg}$, because as seen in Figure \ref{fig:ref-distribution}, it is enough to cover almost all the lengths, similar to a $99\%$ of confidence.

The second step in defining $\delta_i$ is to consider the moving seeds. In the assumed dynamic model, seeds move up to $d_{max}$ discrete units in any direction, randomly chosen with a uniform distribution. Therefore, at any time step, the maximum length of a $R_k$ region is the maximum between $2\times L_{avg}$ or $d_{max}$. This leads to a $\delta_i$ defined as:
\begin{equation}
\label{eq:k_d_def}
    \delta_i = \frac{2^{\big\lceil\log_2(\max(2L_{avg}, d_{max}))\big\rceil}}{2^{i-1}}
\end{equation}
The ceil function is applied on the logarithm because truncated values could lead to an incomplete computation of the VD. Having $\delta_i$ defined, now dJFA works by doing a total of $\log_2(\delta_1)+1$ generational steps. An extra step may be included in order to cover border cases related to the limitations of JFA or the $1\%$ of uncertainty in the distance coverage of $L_{avg}$, that may produce in a few cases an incomplete work. 

The expected behavior of dJFA is that as the seeds set $S$ increases, $\delta_i$ will decrease and this implies fewer steps to be performed. On the other hand, if $S$ is small, then it will behave very much like JFA in terms of performance. Finally if the seed movements are greater than $L_{avg}$, it will trigger the usage of $d_{max}$ as parameter for computing $\delta$, although this is less likely for simulations with smooth moving particles.

\subsection{Combining Moore and Von Neumann Neighborhoods}
\begin{figure}[ht!]
    \centering
    \includegraphics[scale= 0.5]{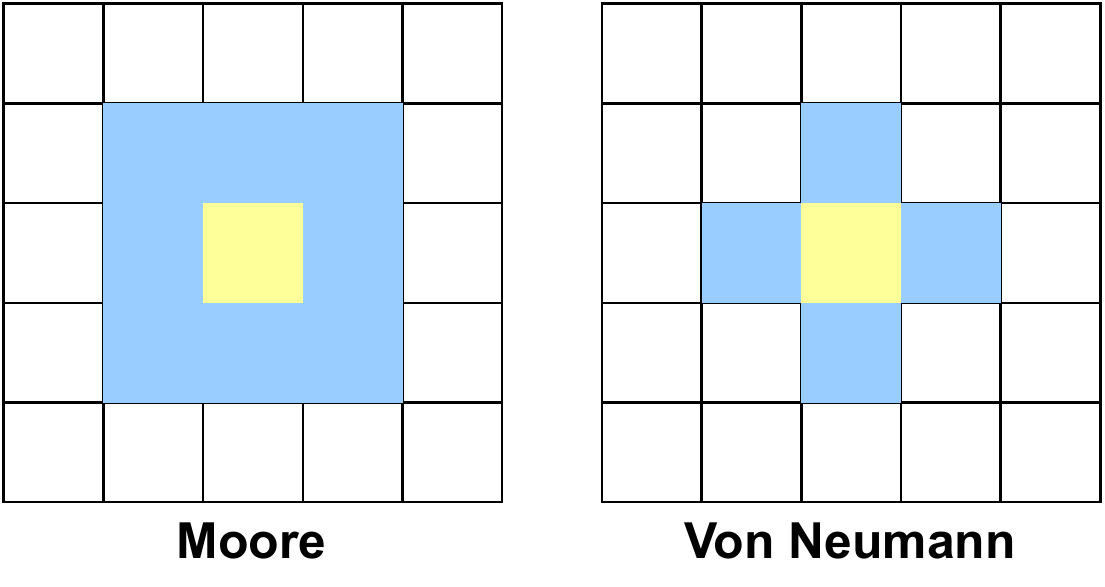}
    \caption{Neighborhoods involved in dJFA.}
    \label{fig:neighborhood}
\end{figure}

In terms of neighborhood, we considered the use of Von Neumann neighborhood instead of the Moore (Figure \ref{fig:neighborhood}). 

\begin{figure}[h!]
    \centering
    \includegraphics[scale=0.10]{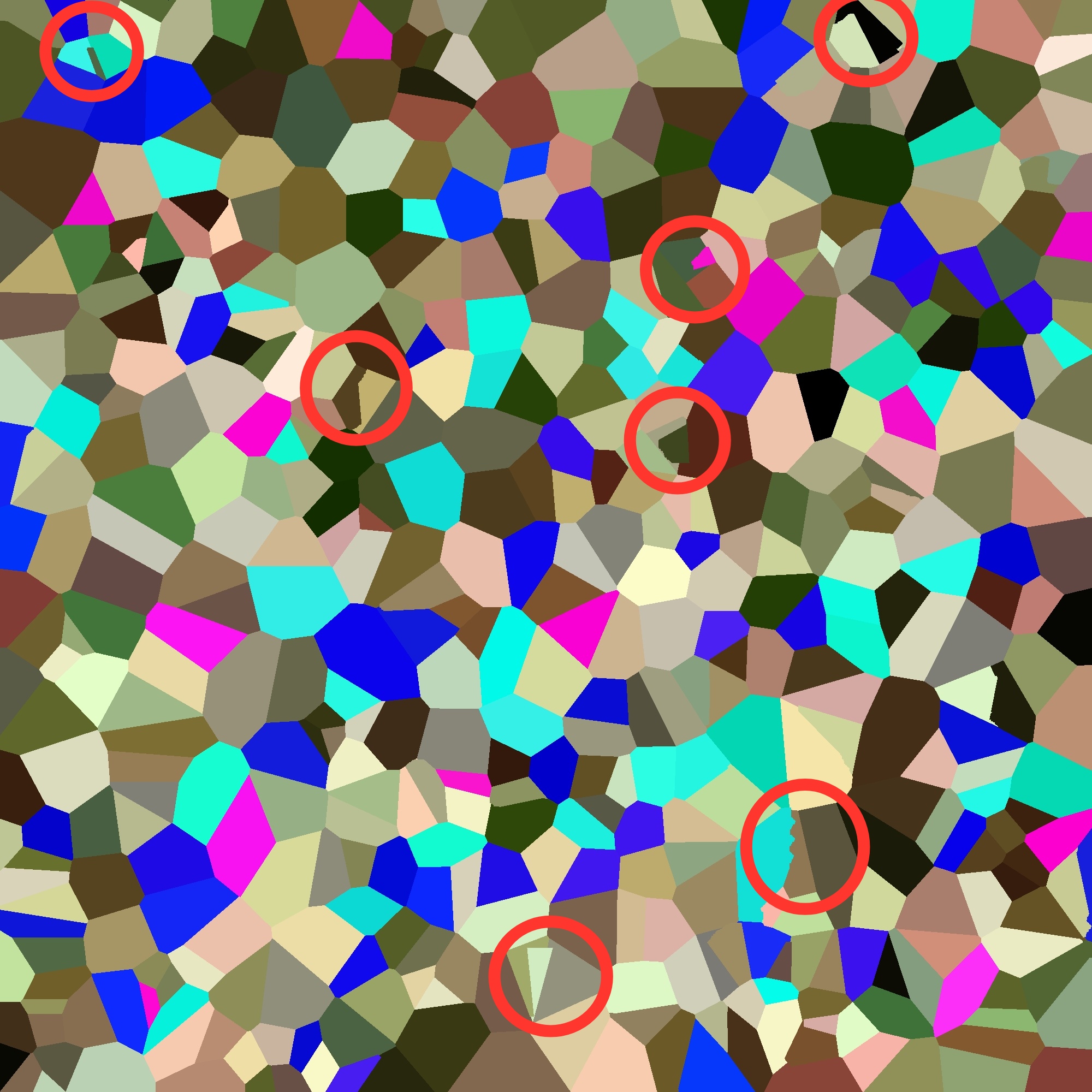}
    \caption{VD built with JFA and only using Von Neumann neighborhood, it can be seen that some regions show some anomalies.}
    \label{fig:bad-vd}
\end{figure}

The motivation to use Von Neumann neighborhood is because it requires exploring half the neighbors compared to Moore, which can speedup the computation although at the cost of generating a less precise VD. Experimental results confirmed that in fact Von Neumann alone generates an incorrect VD even for JFA, as shown in Figure \ref{fig:bad-vd}, where several regions are concave or even generate a saw-tooth border.  
As way to find an intermediate point between performance and correctness, we propose to combine both neighborhoods. The first one or two iterations of dJFA will use the Von Neumann neighborhood, followed by the rest using the Moore one. The reason for this proposal is that in the first iterations many exploration points fall out of the domain or are not the definitive values, therefore a significant amount of work is potentially lost.

\subsection{Euclidean vs Manhattan distances}
The default distance metric is te Euclidean one, but it is also possible to consider the Manhattan distance, which has the advantage of not having square roots neither squared values. We propose two versions of dJFA; i) dJFAe for the euclidean distance version and ii) dJFAm for the Manhattan version. dJFAm is expected to be faster than dJFAe but less precise.

\subsection{dJFA Algorithm Overview}
Algorithm \ref{alg:dJFA} presents the main steps of dJFA working on a generic simulation application with A steps of simulation. 
\begin{algorithm}
    \label{alg:dJFA}
    \caption{dJFA}
\KwData{VD, S, A}
\KwResult{VD}
$k_m \gets ComputeK(\text{VD})$\;
$step \gets 1$\;
\While{$step \le A$}{
    SimulateParticles(S)\;
    $\delta \gets$ computeK(VD,S)\;
    \While{$\delta \ge 1$}{
        \CommentSty{\#Von Neumann|Moore neighborhood\\}
        \textbf{Par}\For{p in VD}{
            \For{q in neighborhood}{
                $s_p \gets VD[p]$\;
                $s_q \gets VD[q]$\;
                \If{$d(s_p,q) < d(s_q,q)$}{
                    $VD[q] = s_p$\;
                }
            }
        }
        $\delta = \delta/2$\;
    }
    $step=step+1$\;
}
\end{algorithm}

The algorithm receives the Voronoi Diagram (VD) grid, the seeds set $S$, and the number of application simulation steps $A$. The outer \texttt{while} loop is for the simulation application. Inside each simulation iteration, a whole dJFA process occurs. First, $\delta$ is computed and then $\log(\delta)$ waves of computation occur. At each wave, parallel GPU threads are launched, mapped to the VD pixels using a one-to-one correspondence. Each thread explores the whole neighborhood, Von Neumann for the first two waves, Moore for the rest. At each neighbor, the thread explores its value and verifies if its propagating seed is closer to the one already assigned. If it is, then such neighbor location is updated with the thread's propagating seed.

\section{Experimental Evaluation}
\label{sec:experimental-evaluation}

\subsection{Experimental Setup}
\begin{figure*}[!ht]
\centering
\includegraphics[scale=0.3]{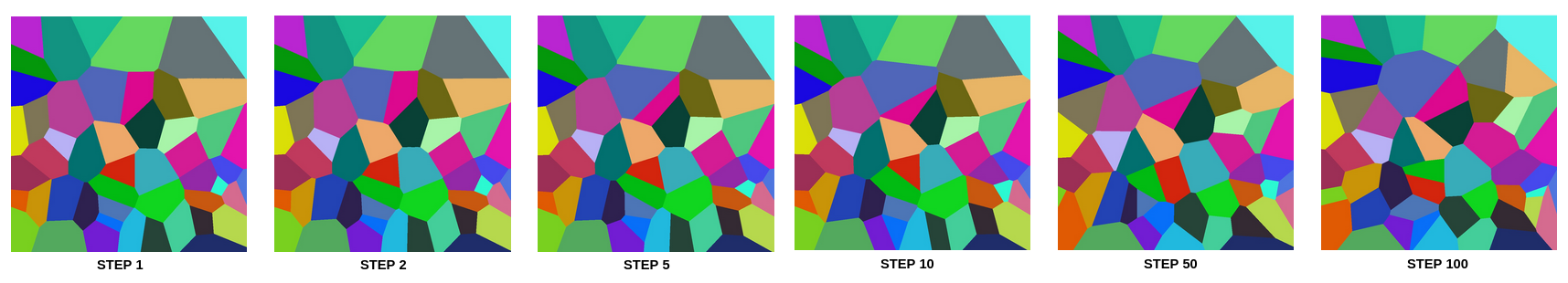}
\caption{A 100 step simulation of 50 seeds on a grid of $1000\times 1000$ pixels. VD samples are obtained at different time steps.}
\label{fig:iterations-vd}
\end{figure*}

The experimental evaluation used one GPU from the Patag\'on Supercomputer \cite{patagon-uach}. Its hardware specifications are listed in Table \ref{tb:specs}.
\begin{table}[!h]
    \caption{Hardware used for tests.}
    \label{tb:specs}
    \centering
    \resizebox{\columnwidth}{!}{
    \begin{tabular}{|c|l|}
        \hline
         System & Patagón Supercomputer - DGXA100 node\\
        \hline
         CPU & $2\times$ AMD Rome 7742 64 cores\\
        \hline
         GPU & $8\times$ NVIDIA A100, 40GB VRAM\\
        \hline 
         RAM & 1TB DDR4\\
         \hline
    \end{tabular}
    }
\end{table}
The benchmark consists on simulating different number of moving seeds on different grid sizes, for $A=100$ application steps using the uniform random distribution model. Figure \ref{fig:iterations-vd} shows how different VDs are obtained throughout the simulation, using a smaller seed count for illustration purposes.

Three metrics are obtained in the simulation benchmark:
\begin{enumerate}
    \item \textbf{Similarity}: is defined in terms of the percentage of equality between dJFA and JFA resulting grids, i.e.,
        \begin{equation}
            \text{Similarity} = 100 \times \frac{\text{matching pixels}}{\text{total pixels}}
        \end{equation}
        
    \item \textbf{Time}: The cumulative time in seconds spent in computing the VD algorithm, ignoring the time spent in moving particles. Times are denoted as $T_{JFA}, T_{dJFAe}, T_{dJFAm}$.
    
    \item \textbf{Speedup}: The acceleration factor defined as
    \begin{equation}
        \text{Speedup} = \frac{T_{JFA}}{T_{dJFA}}.
    \end{equation}
    A second speedup is also considered, $T_{dJFAe}/T_{dJFAm}$, for measuring the acceleration from using the Manhattan distance instead of the Euclidean one.
    
\end{enumerate}

\subsection{Experimental Results}
\begin{figure*}[!htb]
    \centering
    \includegraphics[scale=0.63]{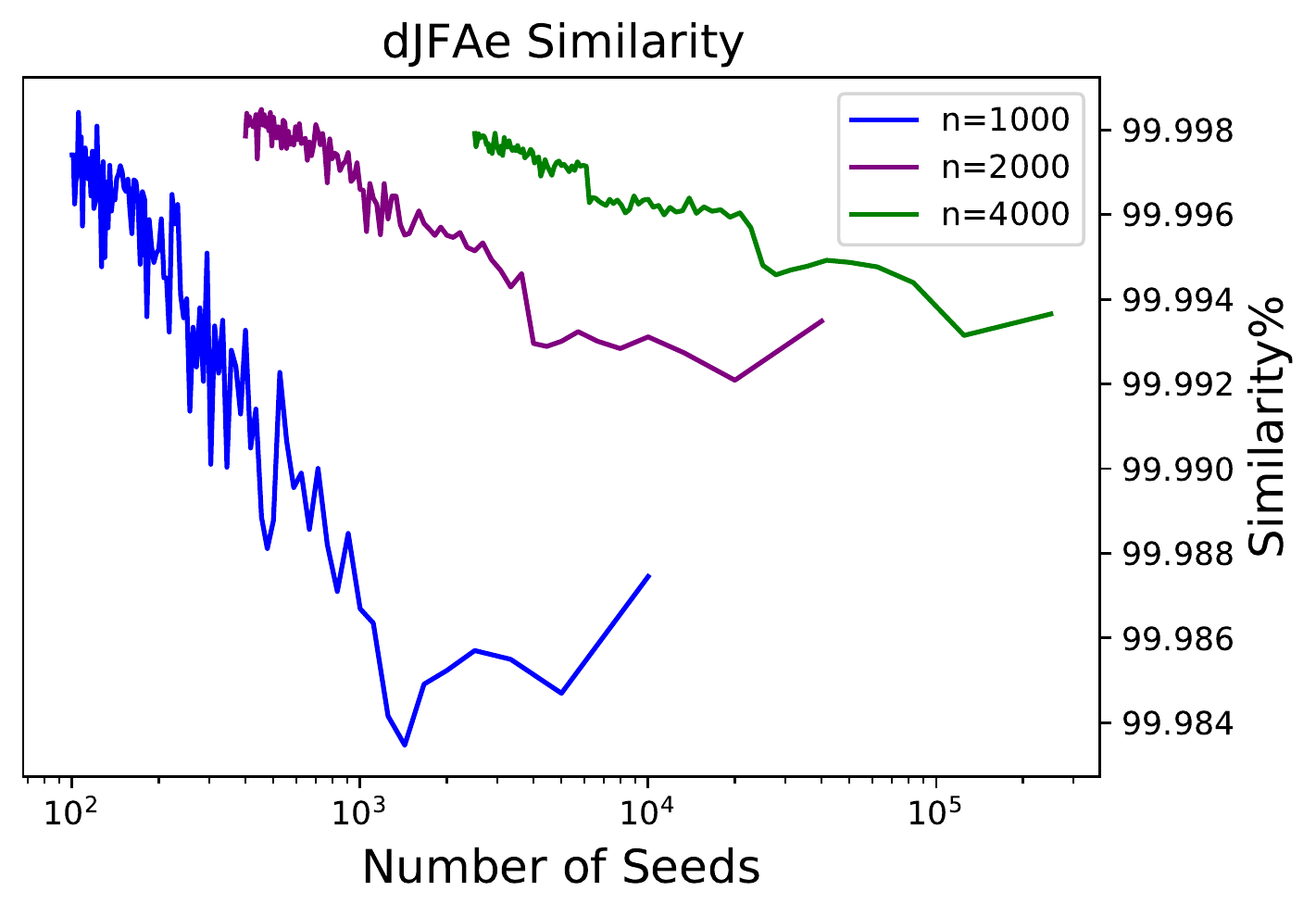}
    \includegraphics[scale=0.63]{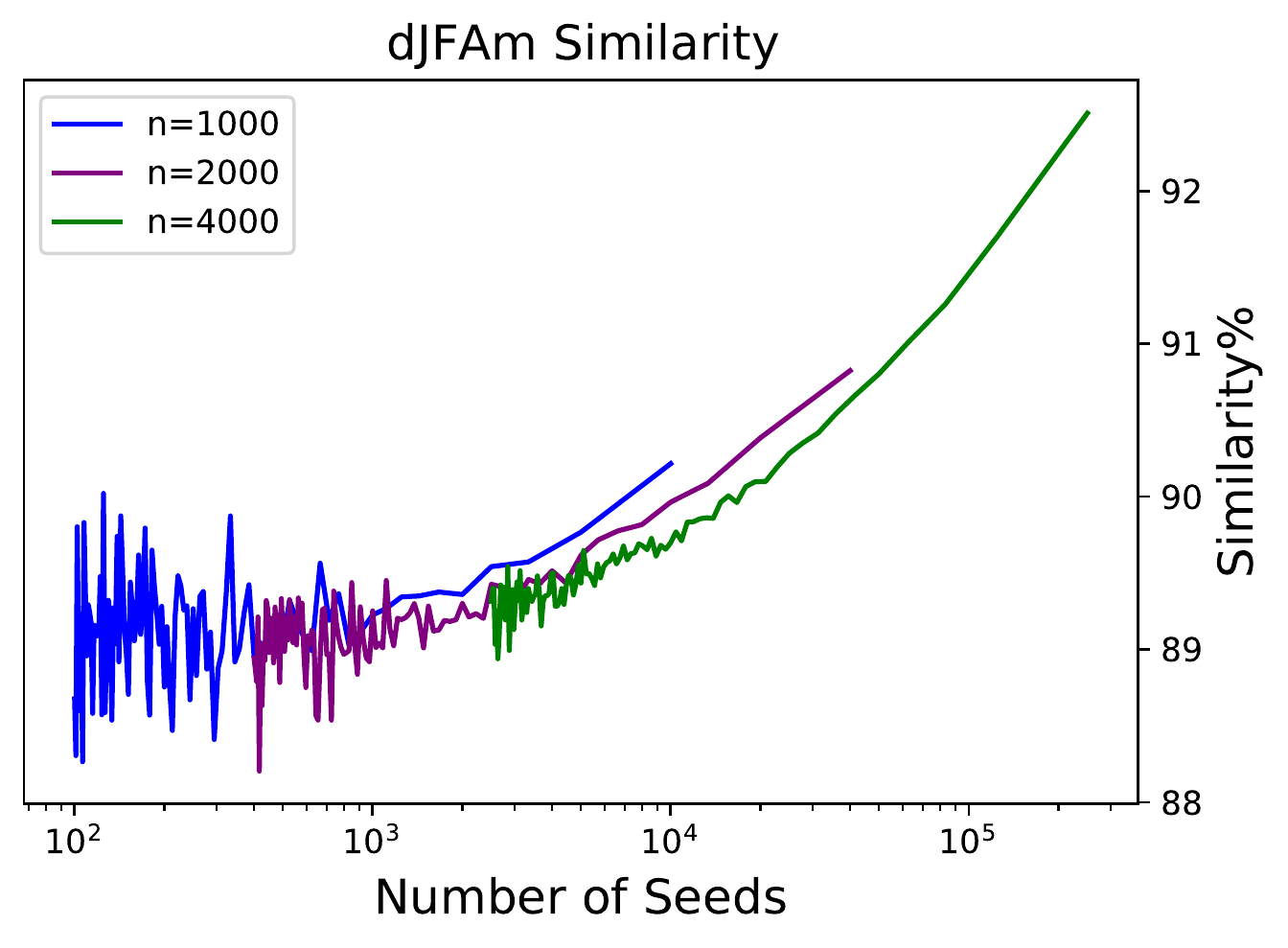}
    \caption{Similarity of dJFAe (left) and dJFAm (right) with respect to JFA.}
    \label{fig:similarity-dJFA}
\end{figure*}

Figure \ref{fig:similarity-dJFA} shows the similarity of both dJFAe and dJFAm with respect to a complete JFA execution. From the plots, one can note how dJFAe is notoriously more precise than dJFAm, reaching nearly $100\%$ of similarity, while dJFAm reaching only $88\%$ to $92\%$ of similarity. It is worth noticing that although dJFAe decreases its similarity as the are more seeds, it shows an stabilization at the end. In the case of dJFAm, one positive aspect is that as the domain is more dense, its similarity increases, each time at a higher rate, making it potentially useful for fully saturated inputs. 

\begin{figure*}[!ht]
\centering
\includegraphics[scale=0.63]{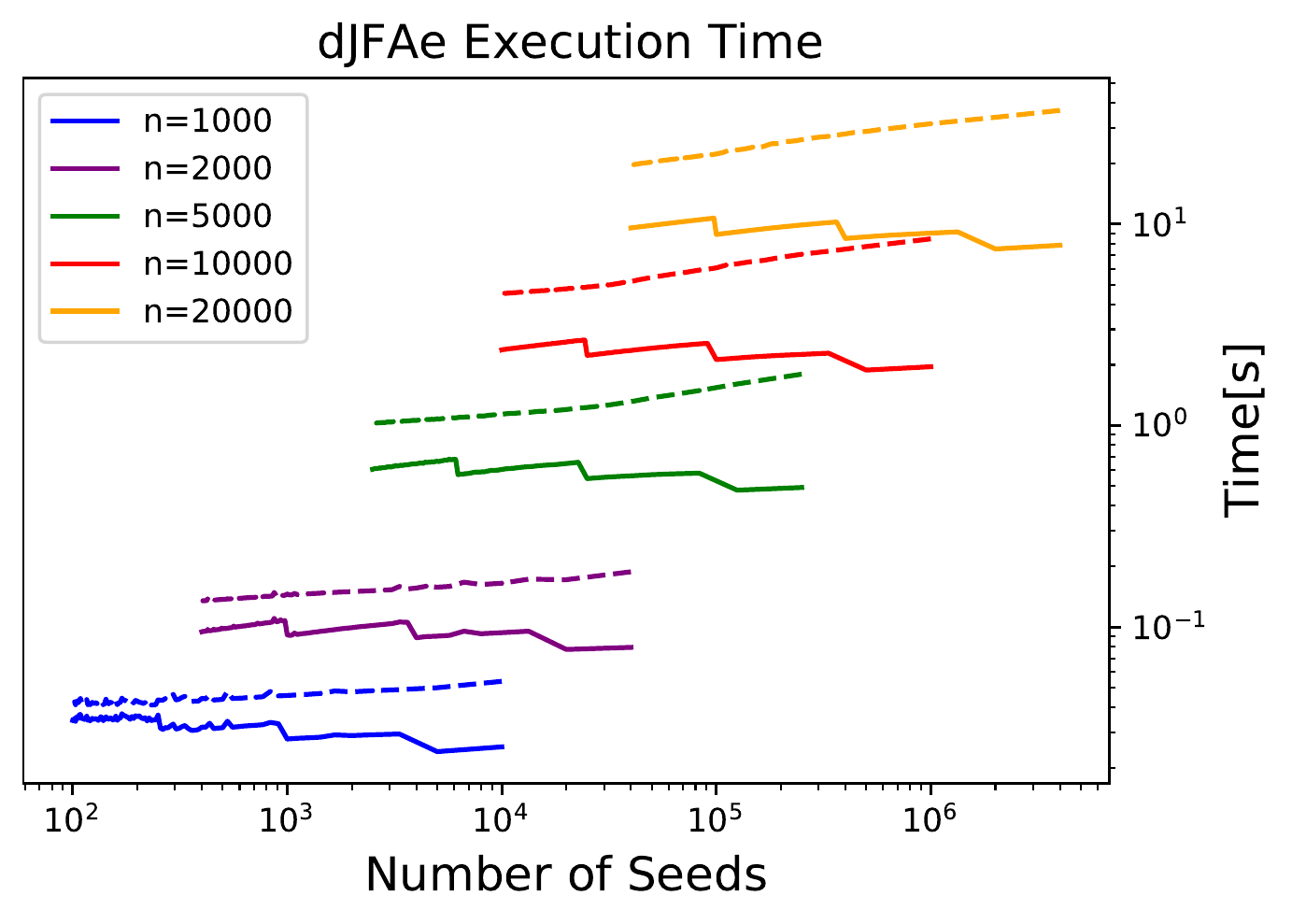}
\includegraphics[scale=0.63]{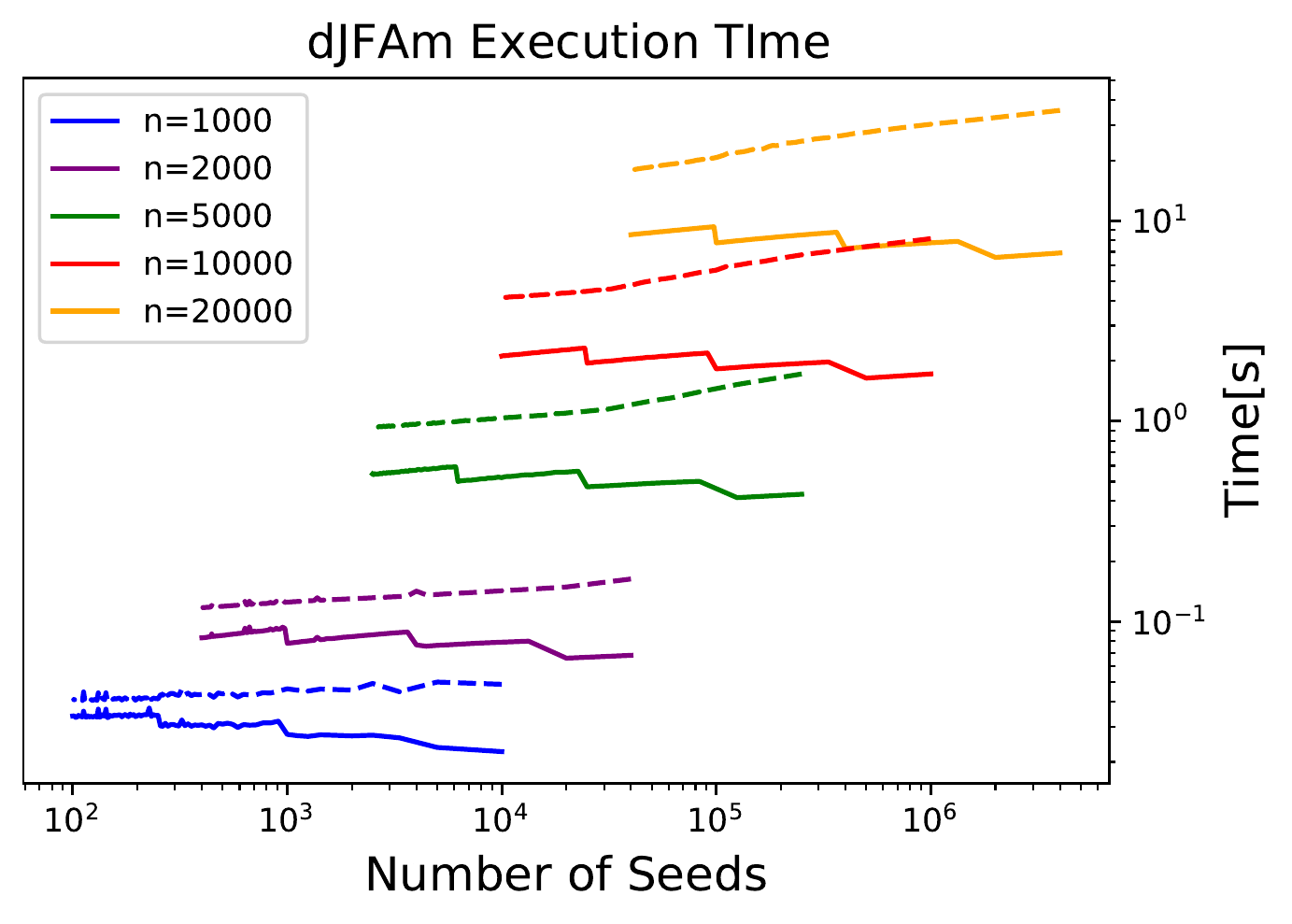}
\caption{Execution times in seconds for JFA (dashed) and dJFAe/dJFAm (solid).}
\label{fig:times-dJFA}
\end{figure*}

The execution times are presented in  Figure \ref{fig:times-dJFA}. 
For all $n$ values, both versions of dJFA took less time than JFA to complete the simulation steps, with the dJFAm version being faster. A staircase pattern can be noticed, where time remains constant until certain values of $n$ are met, where it goes down significantly ($\log$ scale). This pattern is related to the $\delta_i$ definition used, as certain values of $n$, in combination to the number of seeds, make the $\log$ function move to the previous integer value, reducing $\delta_i$ thus the number of iterations and increasing the performance. 

\begin{figure*}[!htb]
\centering
\includegraphics[scale=0.43]{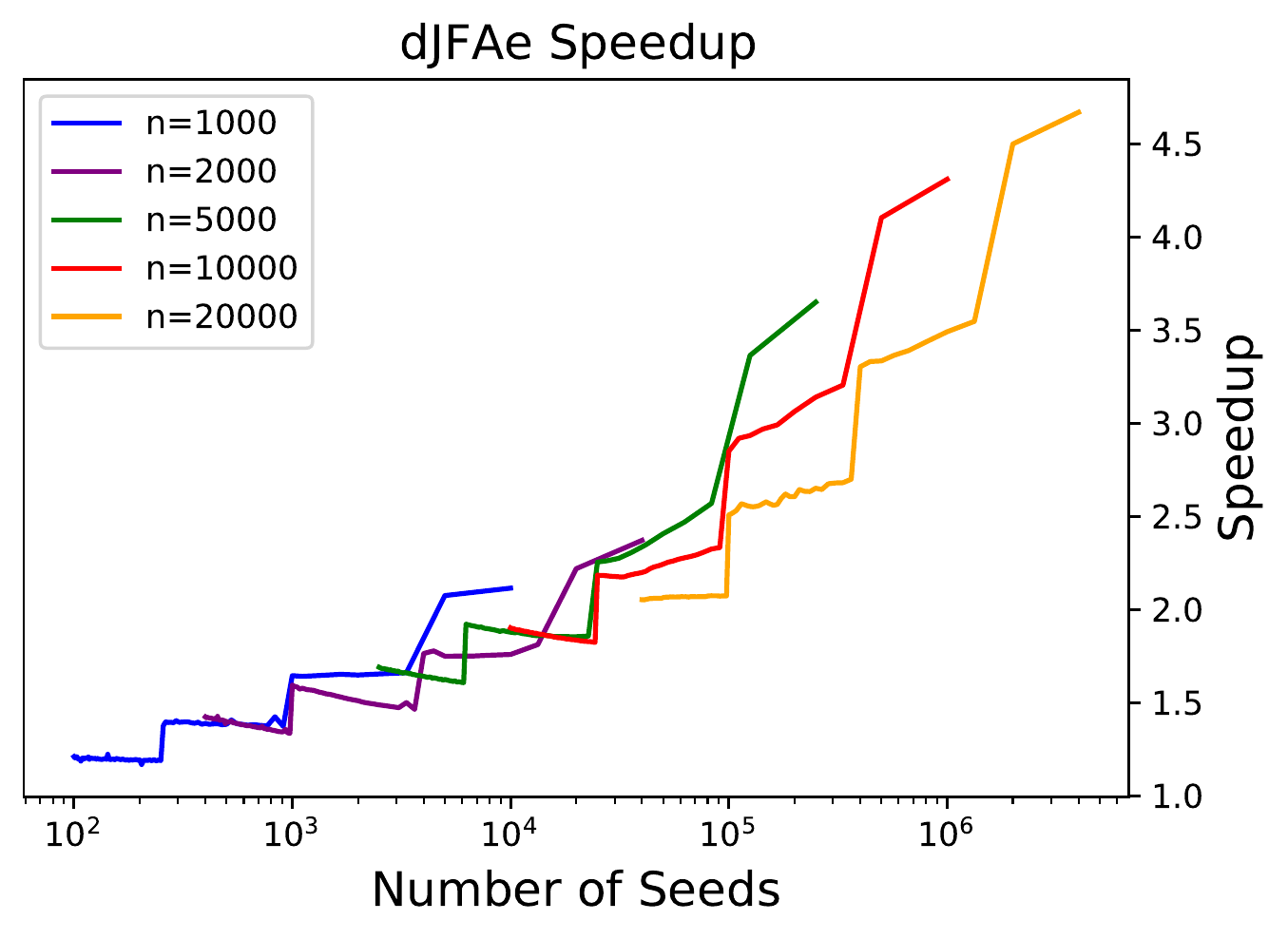}
\includegraphics[scale=0.43]{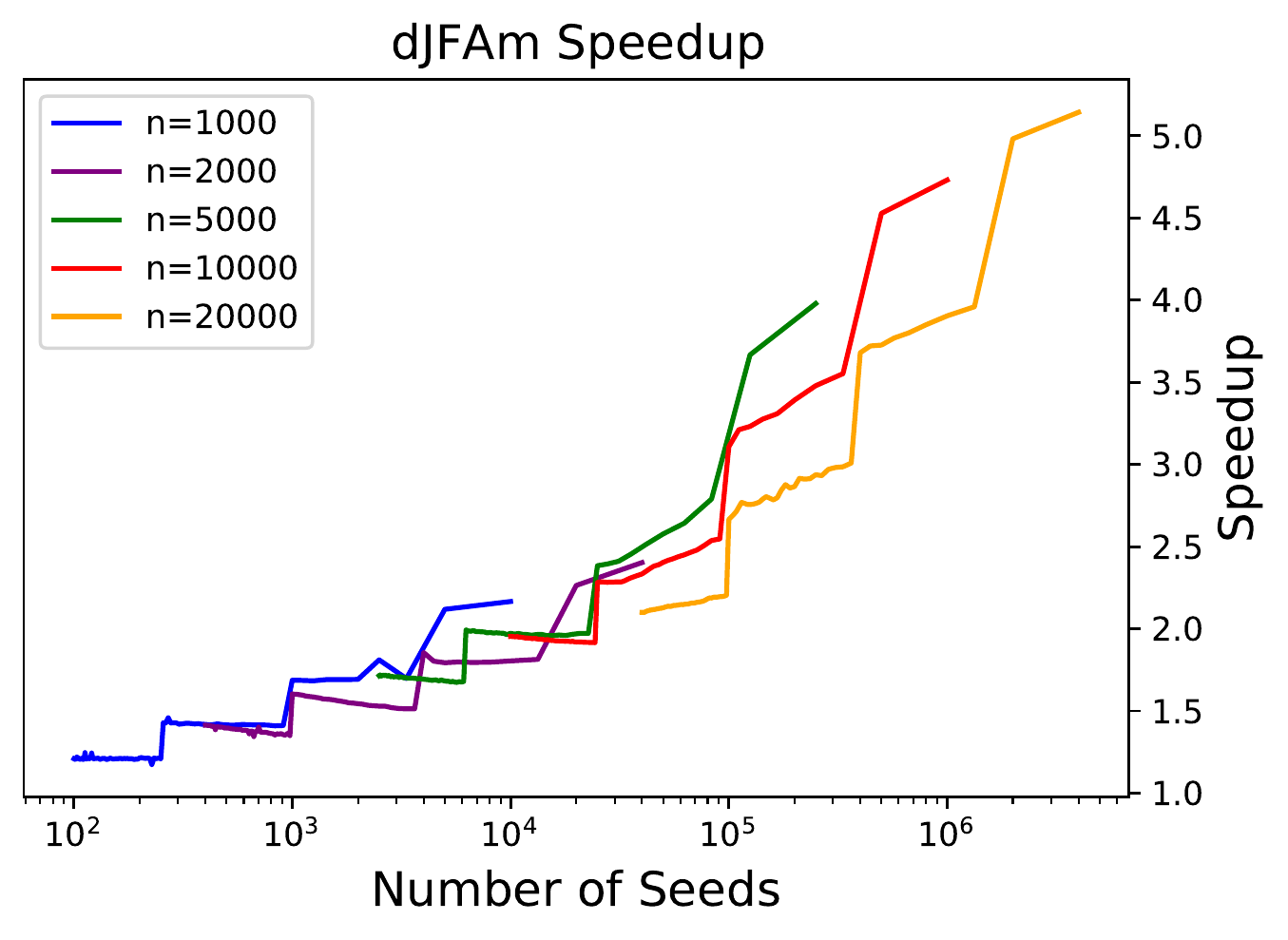}
\includegraphics[scale=0.43]{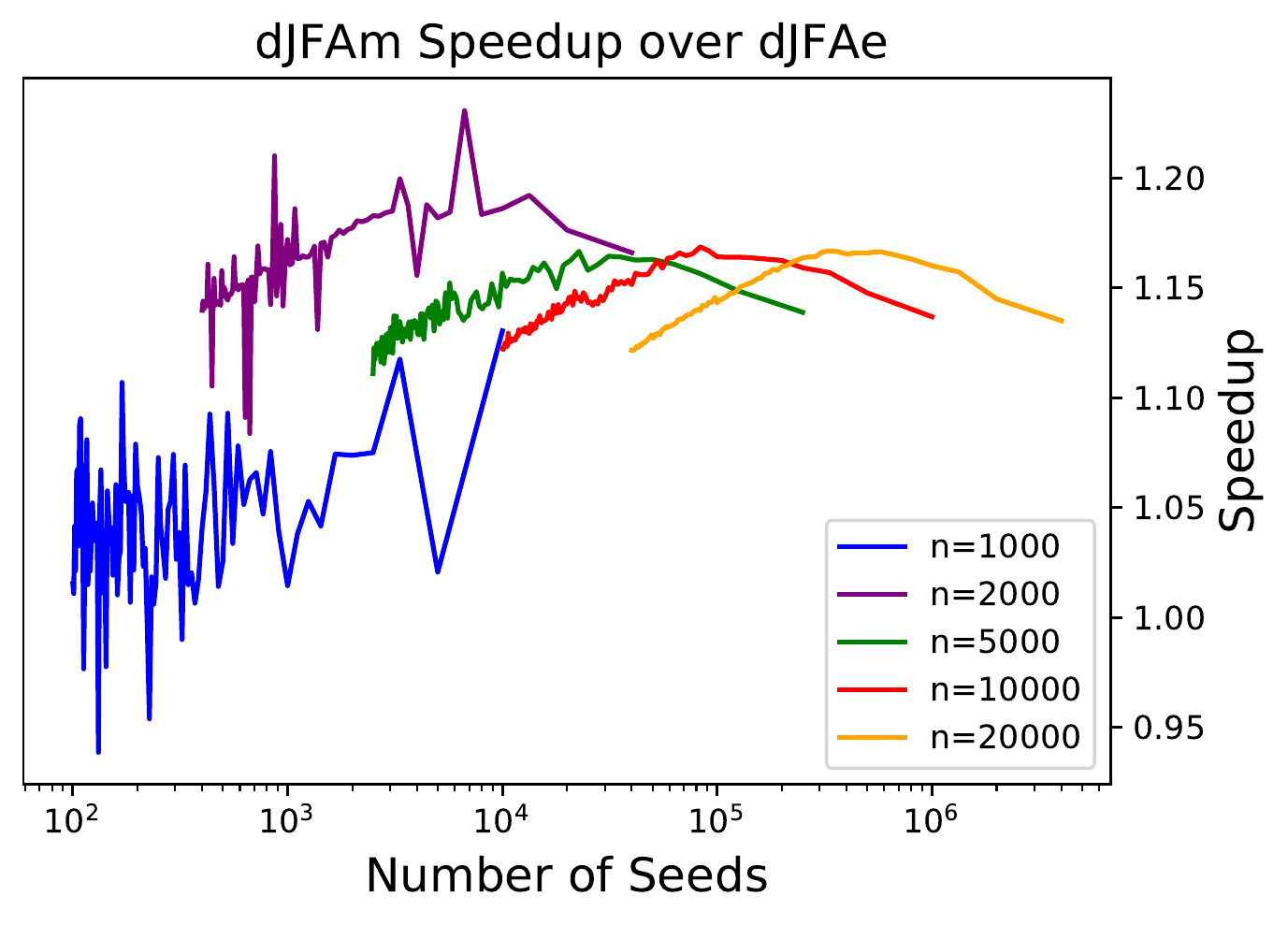}
\caption{Left, speedup of dJFAe (solid) against JFA (dashed). Middle, speedup of dJFAm against JFA. Right, speedup of dJFAm with respect to dJFAm.}
\label{fig:speedup}
\end{figure*}

Speedup results are shown in Figure \ref{fig:speedup}. 
In all tests the speedup is favorable, manifesting the staircase pattern that reaches $\sim 4.5\times$ and $\sim 5\times$ for dJFAe and dJFAm, respectively. When comparing dJFAe with dJFAm, we note that for smaller grids both approaches have relatively similar performance, with dJFAm being slithgly faster. For larger $n$, dJFAm shows a faster performance than dJFAe, with up to $1.2\times$ of speedup in its peak. This is expected as the Manhattan distance has a lesser cost. There is an unexpected decrease in speedup at the end of the plot, meaning that for very dense grids the difference between dJFAe and dJFAm is less relevant. This aspect may require further experimentation and research.

Summarizing what has been observed in all tests, both variants of dJFA manage to achieve better performance than JFA and also with a high degree of similarity depending on the distance function, being the high density scenarios in which the higher speedups were registered. 

\section{Conclusions and Future Work}
\label{sec:conclusions}
This work proposed the dynamic Jump Flooding Algorithm (dJFA), which is an adaptation of the known JFA, now for dynamic moving particles following a uniform distribution. Results show that the proposed method manages to perform faster than a standard JFA. Moreover, the dJFA manages to progressively increase its speedup as the domain gets denser with more seeds. With regard to the similarity, dJFA managed to achieve close to $100\%$ of similarity compared to JFA when using the Euclidean distance metric, otherwise over $~88\%$ of similarity when using the Manhattan distance. This produces to flavors of dJFA, a more precise version with Euclidean distance, or an faster but less precise one with Manhattan distance.

The results also showed aspects in which further work can be done. This work did not consider removal or insertion of seeds during simulation, which is a feature than many applications require. Implementing such feature presents a base challenge of first supporting dynamic arrays in GPU, which is currently a problem under research with some preliminary progress made. Another extension, even more relevant, is to generalize the dJFA to any type of seed movement. This puts a major challenge in the definition of $\delta_i$ as it cannot assume any distribution as this work did. Some ideas include using Dynamic Parallelism or adapt the use of Ray Tracing cores of recent GPUs in order to explore the particles dynamically allowing any initial distribution and movement, even the formation of clusters. Another alternative way to tackle this general case problem is that instead of changing JFA, one can generate a low-resolution version of the grid and do all computation in this reduced space, followed by a reconstruction. Some aspects of this idea have already been developed by the Facet-JFA \cite{10.1145/2683483.2683503}, where low density is exploited to improve the VD computation time. We believe it is possible to take this idea one step further and benefit from recent advances from artificial intelligence, by using Deep Learning Super Sampling (DLSS) to reconstruct the low-resolution results back to the original resolution, which by the way is accelerated by tensor cores. The use and combination of tensor cores with ray tracing cores presents a novel opportunity to keep exploring new possibilities of algorithms for computing dynamic Voronoi Diagrams.

\bibliographystyle{plain}
\bibliography{main} 

\end{document}